\documentclass[twocolumn,showpacs,pre,floatfix,eps]{revtex4}
\ifx\pdfoutput\undefined
\usepackage[demo]{graphicx}
\usepackage{subfig}
\else
\usepackage{epstopdf}
\usepackage{epsfig}
\usepackage{color}

\usepackage{amssymb,amsmath,stmaryrd,tabularx}
\usepackage{wrapfig}
\usepackage{bm}
\fi

\newcommand{\be}{\begin{equation}}
\newcommand{\ee}{\end{equation}}
\newcommand{\bea}{\begin{eqnarray}}
\newcommand{\eea}{\end{eqnarray}}

\newcommand{\ud}{\mathrm{d}}

\newcommand{\ddt}[2]{\frac{\mathrm{d^2}#1}{\mathrm{d}#2^2}}
\newcommand{\ddx}[2]{\ddt{}x}
\newcommand{\ddy}[2]{\ddt{}y}
\newcommand{\ddz}[2]{\ddt{}z}
\newcommand{\pd}[2]{\frac{\partial #1}{\partial #2}}

\renewcommand{\Re}{\operatorname{Re}}
\renewcommand{\Im}{\operatorname{Im}}

\begin{document}
\title{Vortex clustering and universal scaling laws in two-dimensional quantum turbulence}
\author{Audun Skaugen and Luiza Angheluta}
\affiliation{
Department of Physics, \\ University of Oslo, P.O. 1048 Blindern, 0316 Oslo, Norway}


\begin{abstract}
We investigate numerically the statistics of quantized vortices in two-dimensional quantum turbulence using the Gross-Pitaevskii equation. We find that a universal $-5/3$ scaling law in the turbulent energy spectrum is intimately connected with the vortex statistics, such as number fluctuations and vortex velocity, which is also characterized by a similar scaling behavior.  
The $-5/3$ scaling law appearing in the power spectrum of vortex number fluctuations is consistent with the scenario of passive advection of isolated vortices by a turbulent superfluid velocity generated by like-signed vortex clusters. The velocity probability distribution of clustered vortices is also sensitive to spatial configurations, and exhibits a power-law tail distribution with a $-5/3$ exponent.    
\end{abstract}
\maketitle
\date{}

\section{Introduction} 
Richardson's cascade picture of turbulence captures, in essence, the transport of energy across scales in \emph{three-dimensional} (3D) classical fluids: Energy is injected at large scale, and a self-similar breakdown of vortices transports energy down to the dissipative, small scales. This \emph{direct} energy cascade develops because the kinetic energy in 3D turbulence is the dominant statistically invariant quantity, and is associated with Kolmogorov's law for the energy spectrum $E(k)\sim k^{-5/3}$ for wavenumbers $k$ in the inertial range, i.e. in-between the injection and dissipation scales. 

By contrast, \emph{two-dimensional} (2D) classical turbulence manifests itself in a spectacularly different way due to the absence of vortex stretching and twisting. Apart from the kinetic energy, the total vorticity is also a statistical invariant which leads to two inertial cascades. Namely, an~\emph{inverse} energy cascade from small to large scales with the emergence of coherent rotating structures at large scales, and a direct cascade of enstrophy associated with the conservation of vorticity~\citep{robert1980two}. Even though the energy flows in the opposite direction, the turbulent energy spectrum still follows the Kolmogorov $k^{-5/3}$ law in the inverse cascade regime, up to the injection scale below which it crosses over to a $k^{-3}$ scaling associated with the forward enstrophy cascade~\citep{robert1980two}. There is no full consent on the physical mechanism behind the inverse energy cascade. In fact, several physical process have been proposed, such as Kraichnan's picture of~\lq thinning\rq~of small-scale vorticity by strain at large scale~\cite{kraichnan1976eddy,chen2006physical}, and the Onsager's picture of \emph{clustering} of same-signed vortices~\cite{Onsager_1949}.   

Albeit the turbulence phenomenon and its spectral energy transport emerge from the formation and interaction of vortices, the relationship between the statistical properties of turbulence and vortex dynamics is still an open and challenging problem. The role of vortices as the primary structures in turbulence has been long recognized since the pioneering work of Onsager on the statistical description of 2D turbulence in terms of an ensemble of interacting point vortices~\cite{Onsager_1949}. This reduction of turbulence to a complex bundle of vortices has become an effective way of studying turbulence since the realization of turbulent states in quantum fluids~\cite{Paoletti_2011,neely2013characteristics}, and the remarkable discovery that quantum turbulence share similar large-scale statistical properties as classical turbulence~\cite{Paoletti_2011,barenghi2014experimental}. 

Unlike classical vortices which have a diffusive, continuous size and vorticity, quantum vortices are defined by a quantized circulation, i.e. $\Gamma=\oint_C \vec v\cdot \ud \vec l = n\kappa$, where $n$ is an integer and $\kappa=h/m$ is the quantum circulation, which leads to vortex filaments (in 3D) and point-like vortices (in 2D) with well-defined vortex cores. Quantum turbulence is generally referred to as a  complex tangle of these quantized vortices. Despite their quantum nature, the turbulent energy spectrum generated by the interaction of these vortices is characterized by Kolmogorov's classical $k^{-5/3}$ scaling law on scales larger than the mean separation between vortices in superfluids~\cite{Paoletti_2011} and Bose Einstein condensates~\cite{neely2013characteristics}. The similarity between classical and quantum turbulence underscores the universality of turbulence, and the approach to turbulence from vortex dynamics. 

In 3D quantum turbulence, the energy spectrum is associated with a~\emph{direct} energy cascade~\citep{Paoletti_2011,kobayashi2007quantum}. The vortex statistics in the turbulent regime has a particular signature. For instance, vortex line density $L$ is a fluctuating quantity due to vortex interactions. Its frequency power spectrum density decreases as $f^{-5/3}$, which is at odds with the classical interpretation of $L$ as a measure of superfluid vorticity, $\omega = \kappa L$~\cite{Roche_2008}. However, this puzzle was cleared out by a phenomenological model in which the vortex line density $L$ is decomposed into polarized and unpolarized filaments, and the analogy to a passive field advection by turbulence is used to explain the $f^{-5/3}$ spectrum~\cite{Roche_2008,baggaley2011vortex,baggaley2012vortex}. This scaling law of the power spectrum of vortex line fluctuations was observed experimentally in both $^4$He~\citep{Roche_2007} and $^3$He-B~\citep{bradley2008fluctuations}, as well as in numerical simulations of vortex filament model~\cite{baggaley2011vortex,baggaley2012vortex}. In addition to the vortex line statistics, the fluctuations in the superfluid turbulent velocity are also broadly distributed and characterized by a universal $v^{-3}$ power-law tail which has been reported experimentally in superfluids~\citep{Paoletti_2008} and in numerical simulations of 3D trapped Bose-Einstein condensates using the Gross-Pitaevskii equation and the vortex model~\cite{white2010nonclassical}.

While a Kolmogorov $-5/3$ scaling regime in the incompressible kinetic energy spectrum has also been observed in 2D quantum turbulence, the direction of the energy cascade here is more controversial because the compressibility of quantum fluids introduce additional small-scale dissipation by pair-vortex annihilations with phonon emission~\cite{numasato2010direct,Reeves_2012,Reeves_2013,billam2015spectral}. Since it is more tricky to determine the energy flux across scales due to vortex-phonon interactions, the cascade is typically inferred by indirect methods. In systems where energy is injected at long wavelengths and dissipation is most effective at small-scales due to compressibility effects and vortex annihilations, a \emph{direct} energy cascade is proposed to dominate the inertial scaling. This was inferred either from the temporal evolution of the energy spectrum~\cite{numasato2010direct,chesler2013holographic}, or the energy flux across a black-hole event horizon in a holographic gravity dual model of superfluid turbulence~\cite{chesler2013holographic}. On the contrary, an \emph{inverse} energy cascade is attributed to the dynamical regime dominated by vortex clustering and energy injection on scales comparable to the vortex-core size~\cite{Bradley_2012,Reeves_2012,Reeves_2013}. Vortex clustering serve to suppress annihilation events, restoring the conservation of enstrophy in a statistical sense. Several numerical studies of the 2D Gross-Pitaevskii equation and point vortex model~\cite{Bradley_2012,Reeves_2012,Reeves_2013,billam2015spectral} have been focusing on the effect of vortex clustering on the inverse cascade of the incompressible kinetic energy.

The aim of this paper is to study the particular signature of an inverse energy cascade on the statistical properties of vortices. We use the damped Gross-Pitaevskii equation with a stirring potential as proposed in Ref.~\citep{Reeves_2012}, which can simulate a statistically-steady state turbulent regime of a 2D trapped BEC undergoing stirring, where vortices are emitted in clusters in the wake of the stirring obstable. As shown previously~\cite{Reeves_2012}, the incompressible energy spectrum develops a $k^{-5/3}$ power law in the clustering regime. By investigating the effect of vortex clustering on the vortex statistics, we find that the power spectrum of vortex number fluctuations and the distribution of clustered vortex velocities are characterized by a universal power-law behavior with a $-5/3$ exponent as that for the energy spectrum. 

The structure of the paper is as follows. In Section II, we present the damped Gross-Pitaevskii model with a Gaussian stirring potential for simulating 2D turbulence in trapped BEC. The numerical method of tracking vortices and the clustering algorithm for finding clusters of like-signed vortices are detailed in Section III. The incompressible energy spectrum is discussed in Section IV, while the statistics of vortex number fluctuations and vortex velocity fluctuations are presented in Sections V and VI. Finally, Section VII contains a brief summary and concluding remarks.      

\begin{figure}[t]
  \centering
  \includegraphics[width=0.43\textwidth]{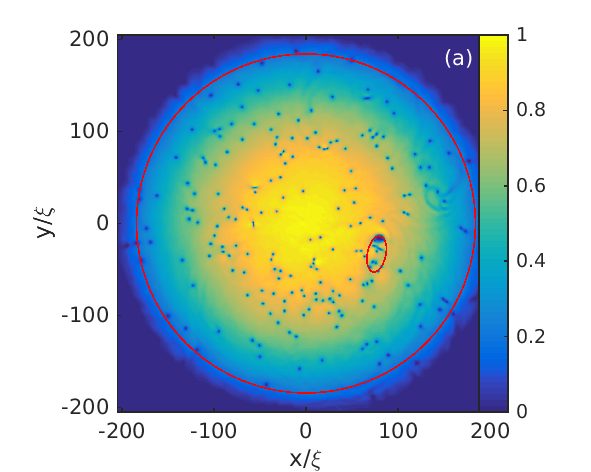}
  \includegraphics[width=0.43\textwidth]{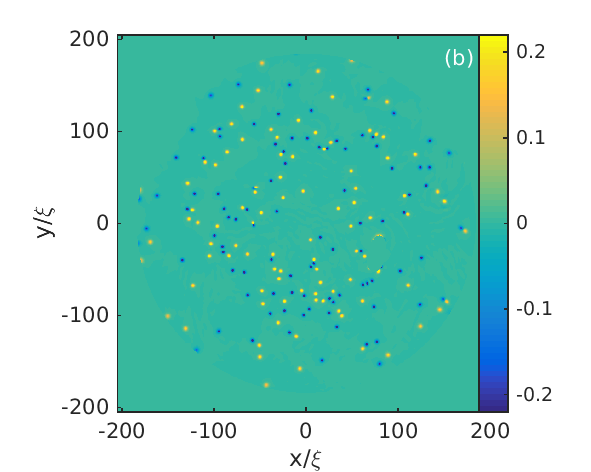}
  \includegraphics[width=0.43\textwidth]{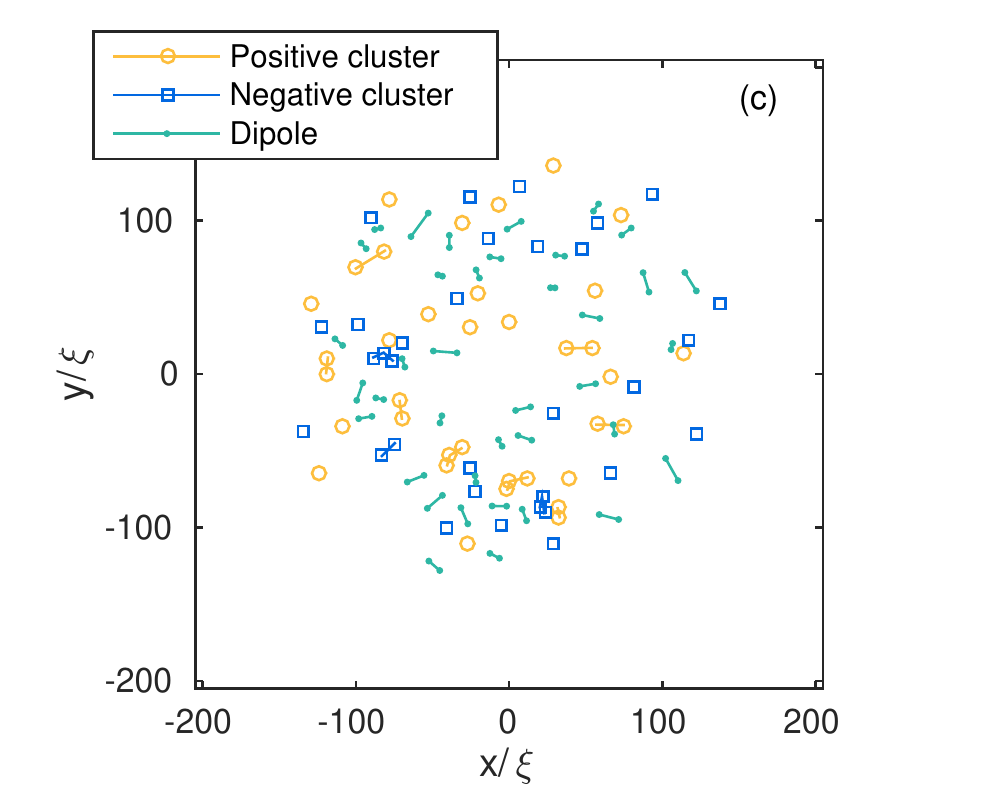}
  \caption{Time snapshot of the density field $|\psi|$ (a), the measure of the quantized vorticity (b), and clustering of the like-signed vortices (c).}
  \label{fig:snaps}
\end{figure}

\section{Gross-Pitaevskii equation} 
We consider a 2D Bose-Einstein condensate described by the wavefunction $\psi(\vec r,t)$, with $|\psi|^2$ related to the particle density of the condensate. The evolution of the wavefunction $\psi(\vec r)$ is described in the mean-field approximation by the damped Gross-Pitaevski equation (dGPE). We use an augmented GPE with a time-dependent external potential for generating statistically-steady state quantum turbulence as proposed in Ref.~\citep{Reeves_2012}. In two dimensions, the dynamics is described by    
\begin{equation}\label{eq:GP}
  \pd{\psi}{t} = (i+\gamma)\left( \frac 1 2 \nabla^2 + 1 - V(\vec r,t) - g|\psi|^2 \right)\psi, 
\end{equation}
with the damping rate $\gamma$ which models phenomenologically the energy dissipation by interaction with a thermal bath. Eq.~(\ref{eq:GP}) is written in dimensionless units by appropriate rescaling of space and time in characteristic units of the coherence length $\xi$ and typical time $\xi/c = \hbar/\mu$, where $c = \sqrt{\mu/m}$ is the speed of sound determined by the chemical potential $\mu$ and the particle's mass $m$. The parameter $g$ describes the nonlinear self-interaction of the condensate, and can be eliminated from the equation by a rescaling of the wavefunction $\psi\rightarrow \sqrt{g}\psi$, thus 
\begin{equation}\label{eq:GP_2}
  \pd{\psi}{t} = (i+\gamma)\left( \frac 1 2 \nabla^2 + 1 - V(\vec r,t) - |\psi|^2 \right)\psi.
\end{equation}
The time-dependent external potential $V(\vec r,t)$ is measured in units of chemical potential $\mu$, and consists of a trapping potential $V_t(\vec r)$ and a time-dependent stirring potential $V_{\text{ext}}(\vec r,t)$ used to generate quantum turbulence as proposed in Ref.~\cite{Reeves_2012}. The trap is given by the harmonic potential as
\begin{equation}
V_t(\vec r) = \frac 1 2 \omega_t^2 r^2, 
\end{equation}
such that it causes the Thomas-Fermi solution of $\psi$ to 
vanish when the harmonic potential exceeds the chemical potential, thus for radii larger than the Thomas-Fermi length $R_{TF}= \sqrt{2}/\omega_t$. Given a desired size of $R_{TF}$ of the condensate, we can therefore choose the parameter as $\omega_t = \sqrt 2/R_{TF}$. 

The time-dependent stirring potential $V_{\text{ext}}(\vec r, t)$ is given by a Gaussian obstacle centered at $\vec r_{\text{ext}}(t)$,
\begin{equation}
  V_{\text{ext}}(\vec r, t) = 	V_0 \exp\left \{ -\frac{|\vec r - \vec r_{\text{ext}}(t)|^2}{w^2} \right\},
\end{equation}
where $V_0>1$ is the height of the obstacle, and the width is set to $w = 4\xi$. The center of the obstacle moves in a circle of radius $R_{\text{ext}} = 0.4 R_{TF}$ with a speed $v_{\text{ext}}$, so that 
\begin{equation}
  \vec r_{\text{ext}}(t) = R_{\text{ext}}\left[ \cos \left(\frac{v_{\text{ext}}}{R_{\text{ext}}}t\right) \vec i+  \sin\left(\frac{v_{\text{ext}}}{R_{\text{ext}}}t\right) \vec j \right],
\end{equation}
where $\vec i$ and $\vec j$ are unit vectors in the $x-y$ plane. 

We set the values of the model parameters in the parameter space associated with the turbulent regime~\citep{Reeves_2012}. We consider the particular values $V_0 = 1.4\mu$, $R_{TF}= 0.8\times 256 \xi$ and $\gamma=0.009$, and vary the stirring speed $v_{\text{ext}}$ of the Gaussian obstacle so that we obtain a more robust clustering of vortices in the wake of the obstacle.

We solve the dGPE from Eq.~(\ref{eq:GP_2}) numerically by using spectral methods with exponential time differencing, and study different dynamical regimes depending on the different values of the stirring speed $v_{\text{ext}}$. 

\begin{figure}[t]
  \centering
 \includegraphics[width=0.45\textwidth]{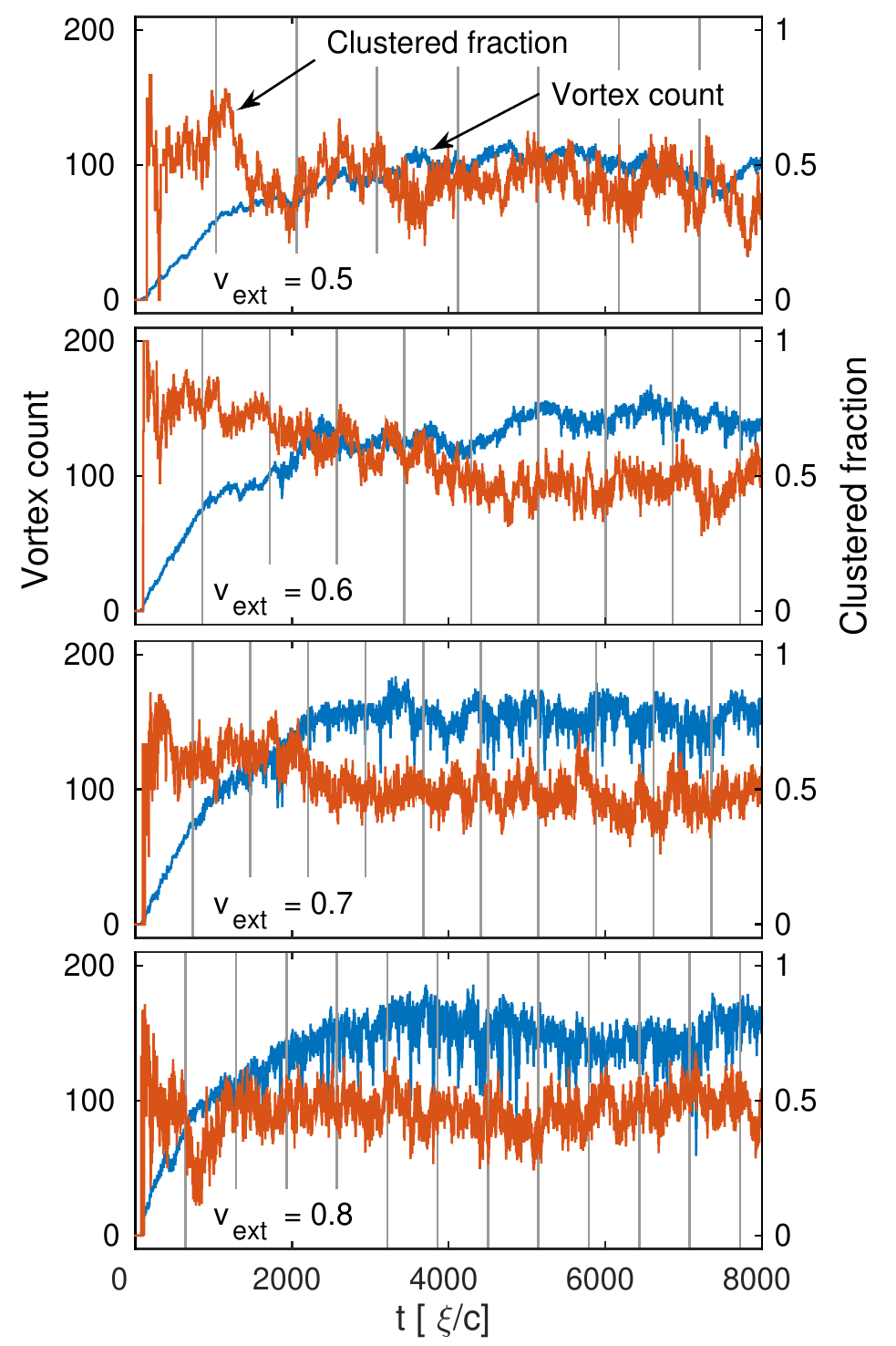}
  \caption{Time windows of the vortex density fluctuations versus fraction of vortex clusters for different values of the stirring velocity $v_{\text{ext}}$. 
Vertical lines mark times where the stirrer has moved in a complete circle around the condensate.}
  \label{fig:cfracs}
\end{figure}

\section{Tracking and clustering of vortices} 
\label{sec:track}
We locate the position and velocity of quantized vortices directly from the wavefuntion $\psi$ using the field formulation of Halperin~\cite{halperin1981statistical} and Mazenko~\cite{mazenko2001}. A similar numerical implementation of this method was studied numerically for Ginzburg-Landau and Swift-Hohenberg dynamics in Ref.~\cite{Angheluta_2012}. The key insight behind this method is that vortices occur exactly where the wave function vanishes inside the Thomas-Fermi radius, i.e. for $V < \mu$. The zeroes of the $\psi(\vec r,t)$-field can be related to the density of vortices $\rho(\vec r,t)$ by the transformation $\rho(\vec r,t) = \delta(\psi)\mathcal{D}(\vec r,t)$, with the Jacobian determinant given by~\cite{mazenko2001,Angheluta_2012}
\begin{eqnarray}
\mathcal{D} = \left| \begin{array}{cc} \partial_x \Re\psi & \partial_y\Re\psi \\ 
\partial_x\Im\psi &\partial_y\Im\psi\end{array} \right | 
= \Im\left( \partial_x\psi^*\partial_y\psi \right).
\end{eqnarray}
This Jacobian field is zero everywhere except within a vortex core, and its sign indicates the rotational direction of the vortex. We can therefore apply a threshold to the determinant in order to locate these vortex cores. However, due to the presence of the external potential $V$, the maximum value of $\mathcal{D}$ will vary roughly as $|\psi_{TF}|^2 = 1-V$. We therefore normalize the field $\mathcal{D}$ by this factor, and search for regions where $\mathcal{D}/(1-V)$ exceeds a given threshold. There are two spatial regions where this method becomes inapplicable. One is the boundary region close to $R_{TF}$, where the wavefunction vanishes quickly. The other one is the stirring obstacle and its wake where a dense collection of vortices are frequently nucleated and vortex cores might not be isolated from each other. To remove these boundary effects, we apply two masks to the normalized $\mathcal{D}$-field before applying the threshold. The boundaries of these masks are drawn on the absolute value of the wavefunction in Figure \ref{fig:snaps} (a), and correspond to setting to zero the value of $\mathcal{D}$ outside a circle of radius $0.9R_{TF}$ and inside an ellipse with the stirrer at one focus. The resulting normalized $\mathcal{D}$-field with these boundary masks applied is shown in Figure~\ref{fig:snaps}(b).

Vortex positions can now be located by calculating the center of mass of each connected region found. As discussed in the introduction, an inverse energy cascade is associated with clustering of vortices of the same circulation. We therefore implement a clustering algorithm using the method outlined in Ref.~\cite{Reeves_2013}. In this algorithm, a pair of oppositely charged vortices are considered a dipole if they are closer to each other than either is to any other vortex. Two like-charged vortices are considered part of the same cluster if they are closer to each other than either is to any oppositely charged non-dipole vortex.
The resulting vortex positions and the clusters of like-signed vortices are shown in Figure~\ref{fig:snaps}(c).

The clustering analysis allows us to measure the \emph{clustered fraction}, defined as the number of \emph{clustered} vortices relative to the total number of vortices. The fluctuations in the clustered fraction is compared with those in the number of vortices as shown
in Figure~\ref{fig:cfracs} for different values of the stirring velocity. We notice that the vortex count is increasing from zero and then is fluctuating around a steady-state value after few rounds of the stirring obstacle. There is an initial spike in clustering, as the stirring obstacle readily creates clusters. This high amount of clustering is however not sustainable, and the clustered fraction settles at a lower level. The vortex counts fluctuate very little in the initial stages, when the amount of clustering is large. This makes sense because clustered vortices seldom interact with opposite-signed vortices to annihilate with. As the amount of clustering settles down, the fluctuations increase in strength. We also observe that there is a tendency for the clustered fraction to fluctuate towards larger values when the stirring obstacle moves slower.

\begin{figure}[t]
  \centering
  \includegraphics[width=0.45\textwidth]{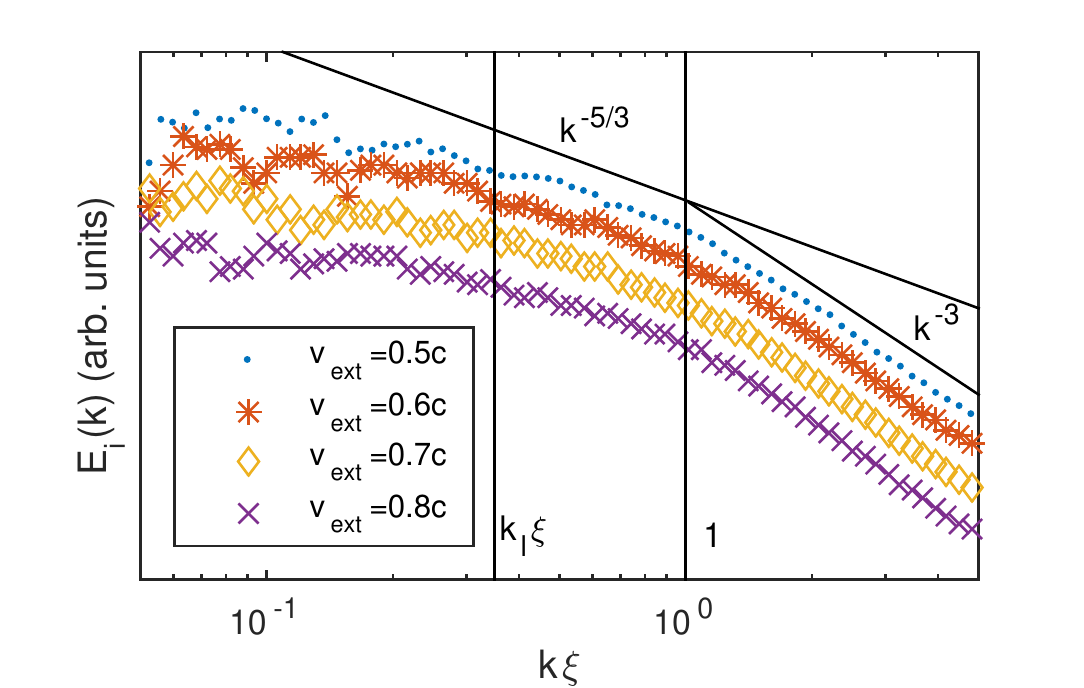}
  \caption{Loglog plot of the incompressible energy spectra at $t = 5060\xi/c$ for the various stirring velocities. The two vertical lines mark $k\xi = 1$, where the core structure becomes important, and the approximate wavenumber 
  $k_l = 2\pi/l$ corresponding to the mean intervortex distance $l$. The different spectra are shifted vertically for comparison.}
  \label{fig:especs}
\end{figure}

\begin{figure}[t]
  \centering
  \includegraphics[width=0.45\textwidth]{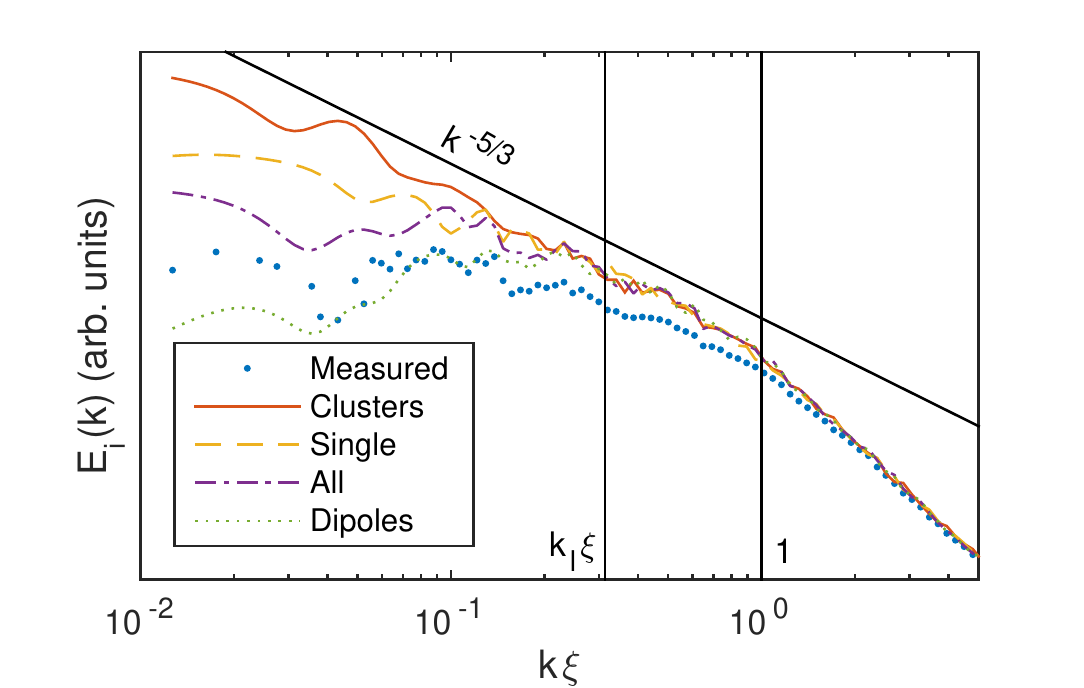}
  \caption{Analytical incompressible energy spectra calculated from the vortex configuration found at $t = 5060\xi/c$ with $v_{\text{ext}} = 0.5c$. This is compared to the numerical energy spectrum directly calculated from the velocity field, as described in Section~\ref{sec:especs}.}
  \label{fig:synenspecs}
\end{figure}

\section{Energy spectrum}
\label{sec:especs}
Quantum turbulence in BEC is characterized by a cascade of energy across inertial scales analogous to that of a turbulent flow in classical fluids. An energy cascade is associated with a kinetic energy spectrum which exhibits a $-5/3$ power-law over the inertial wavenumbers. The classical energy spectrum is obtained by a spectral decomposition of the kinetic energy of an incompressible fluid, 
\begin{equation}
E_{kinetic} = \frac 1 2 \int\ud^2 \vec r \rho v^2 = \int\ud k E(k), 
\end{equation}
where the classical fluid density $\rho$ is constant, and the energy spectrum is the accumulated energy in a shell in the $\vec k$-space,
\begin{equation}
E(k) = \int_{|\vec k| = k}\ud^2 \vec k\mathcal{E}\left(\vec k\right) . 
\end{equation}
Using homogeneity and the convolution theorem, this spectral density can be calculated from the Fourier-transformed velocities as 
\begin{equation}
\mathcal{E}(\vec k) = \frac 1 2 \rho \vec v\left(\vec k\right)\cdot \vec v\left(-\vec k\right).
\end{equation}
By analogy, the same definition of the energy spectrum for the BEC holds, but with two modifications due to the compressibility of a quantum fluid~\citep{Nore_1997}. Firstly, as the density of the superfluid is not constant, the superfluid velocity field obtained by the Madelung transformation must be weighted by the square-root of the density as  
\begin{equation}
\vec u(\vec r) = \sqrt{\rho(\vec r)}\nabla\theta(\vec r), 
\end{equation}
where $\theta(\vec r)$ is the phase of the wave function $\psi$. Secondly, this weighted velocity field is decomposed into compressible and incompressible parts, as follows
\begin{eqnarray}
\vec u = \vec u_c + \vec u_i\textrm{, where }\nabla \cdot \vec u_i = 0\textrm{ and } \nabla \times \vec u_c = 0. 
\end{eqnarray}
The incompressible energy spectrum is then calculated as
\begin{equation}\label{eq:Ei}
E_i(k) = \int_{|\vec k| = k}\ud^2 \vec k \left[\vec u_i\left(\vec k\right)\cdot \vec u_i\left(-\vec k\right)\right]. 
\end{equation}
We calculate the incompressible energy spectrum from Eq.~(\ref{eq:Ei}) for different speeds of the stirring obstacle and the result is shown in Figure~\ref{fig:especs}. We notice that at wavenumbers larger than $1/\xi$ (corresponding to scales smaller than $2\pi\xi$), the energy spectrum follows a universal $k^{-3}$ power-law tail independent of the stirring velocity and the model parameters. Although the $-3$ exponent is the same as that for the enstrophy cascade in two-dimensional classical turbulence, in the case of quantum turbulence this regime is determined by the quantum vortex core structure~\citep{Bradley_2012}. The energy injection scales falls in the intermediate scales between the vortex core size and the mean vortex separation $l$, where the $k^{-3}$ scaling also crosses over to a different regime. On lengthscales larger than $l$ but smaller than the Thomas-Fermi radius, equivalently for $2\pi/R_{FT}<k<2\pi/l$, a $k^{-5/3}$ starts to develop in association to vortex clustering. Admittedly, this wavenumber range is too narrow to confidently claim the existence of an inertial scaling regime, although it was assumed in previous similar studies~\citep{Reeves_2012,Reeves_2013}. In fact, it was recently shown in Ref.~\cite{billam2015spectral} that an accidental $k^{-5/3}$ may occur as a cross-over regime between the two asymptotic scaling regime of the energy spectrum of isolated vortices, i.e. $k^{-1}$ at large $k$'s and $k^{-3}$ at small $k$'s, and that it disappears when the effect of vortex-core size is removed. To test that the $k^{-5/3}$ is indeed a true scaling regime, but very limited by finite size effects and an insufficient separation of scales, we seek to control these effects by separating out the contribution due to vortex clusters.

\subsection*{Energy spectrum of clustered vortices}
One possible reason for the poor Kolmogorov scaling signal in Figure~\ref{fig:especs} is the fact that there are many more isolated vortices and dipoles compared to vortex clusters, as we can see in Figure~\ref{fig:cfracs}. Another important effect that we believe may dominate the statistics is the limited separation of scales between the Thomas-Fermi radius and the mean vortex separation. 

As discussed in previous works, e.g.~\cite{Bradley_2012,Reeves_2012,Reeves_2013,billam2015spectral}, the inverse energy cascade in two dimensional quantum turbulence is attributed to clustering of like-signed vortices. In order to isolate the contributions of clustered vortices to the energy spectra, we use the analytical approach from Ref.~\cite{Bradley_2012} to calculate the energy spectrum resulting from a given configuration of vortices taken from our numerical simulations of the dGPE. 

Based on the superposition principle of the velocity induced by $N$ well-separated vortices, the incompressible energy spectrum can be determined by the energy spectrum of a single vortex and the configurational distribution function of vortices as~\cite{Bradley_2012}
\begin{equation}\label{eq:Ei_N}
E_i^N(k)\propto F_\Lambda(k\xi)G_N(k),
\end{equation} 
where $F_\Lambda(k\xi)$ is the single-vortex energy spectrum calculated as 
\begin{equation}
F_\Lambda(k\xi) = \Lambda^{-1}f(k\xi \Lambda^{-1}),
\end{equation} 
with $\Lambda$ being the slope of the wavefunction of the center of the vortex core and 
\begin{equation}
f(z) = \frac{z}{4}\left[ I_1\left(\frac{z}{2}\right)K_0\left(\frac{z}{2}\right)-I_0\left(\frac{z}{2}\right)K_1\left(\frac{z}{2}\right)\right]^2.
\end{equation} 
The configurational function $G_N(k)$ for $N$ vortices with positions $\vec r_p$ and circulation signs $\kappa_p=\pm 1$ is calculated as~\cite{Bradley_2012} 
\begin{equation}
G(k) = 1+\frac{2}{N}\sum_{p=1}^{N-1}\sum_{q=p+1}^{N}\kappa_p\kappa_q J_0\left(k|\vec r_p-\vec r_q|\right).
\end{equation} 
We use the clustering analysis described in the previous section to extract the position of clustered vortices from vortex configurations obtained numerically. We then calculate the separate contribution to the energy spectrum of various subsets of vortices using Eq.~(\ref{eq:Ei_N}). The result is illustrated in Figure~\ref{fig:synenspecs}. The energy spectrum due to all vortices follows the measured energy spectrum closely, apart from lower measured energies at low wavenumbers. This is because the analytical solution does not take into account the density profile of the superfluid, which drops off to lower values at larger scales. If we isolate the contribution of vortex clusters, the Kolmogorov $k^{-5/3}$ scaling laws extends to much smaller wavenumbers, and approaches a decade of scaling. Also, we have checked that this scaling law persists for $k<2\pi/l$ even when we remove the effect of the core size. 

\section{Vortex number fluctuations}
To understand the connection between the statistical properties of turbulence, e.g. energy spectrum, and vortex dynamics, we study the statistics of vortex number fluctuations and vortex velocity.  

\begin{figure}[t]
  \centering
  \includegraphics[width=0.45\textwidth]{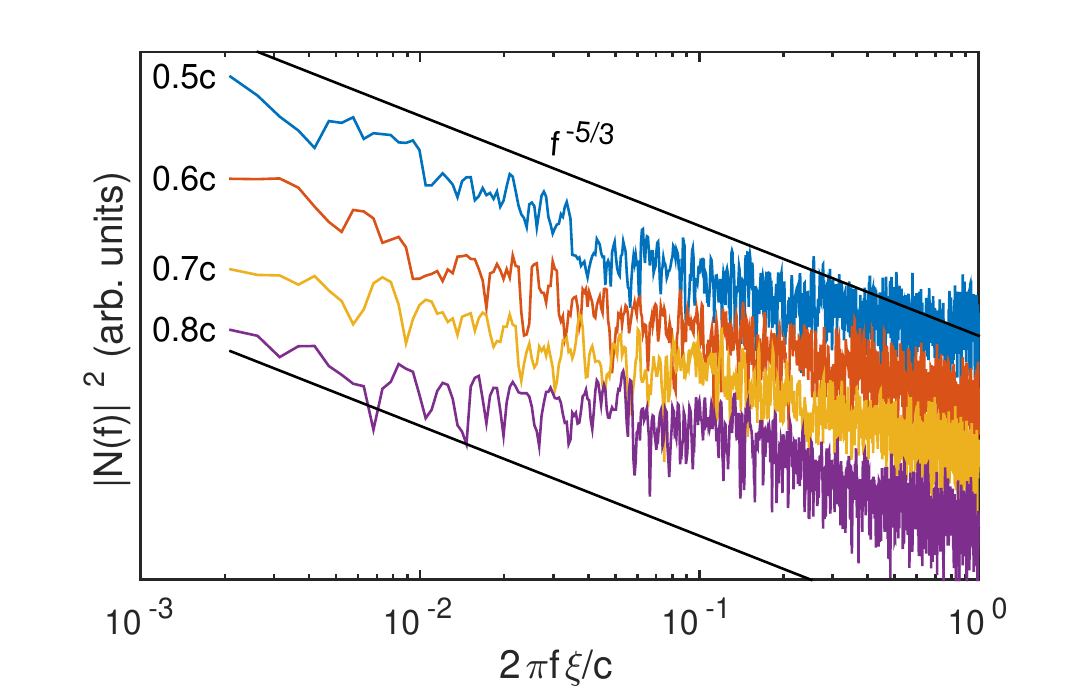}
  \caption{Loglog plots of the power spectra of vortex number fluctuations, at different stirring velocities. The spectra are shifted vertically to keep them from overlapping.}
  \label{fig:pss}
\end{figure}

We investigate the effect of the vortex clustering on vortex number fluctuations, in terms of their power spectrum. Temporal fluctuations of vortex counts are marked by a transient period due to the nucleation of vortices in the wake of the stirring obstacle. This transient time is excluded from the statistics, therefore we look at fluctuations in the steady-state regime, i.e. $t > 4000\xi/c$. The resulting power spectra are shown in Figure \ref{fig:pss}. We see a power-law decay with an exponent close to $-5/3$, at least for the lower values of the stirring velocities.

The same power-law exponent for the power spectra of the vortex line density was reported experimentally and numerically for 3D counterflow turbulence in the superfluid helium~\citep{Roche_2007,baggaley2011vortex,baggaley2012vortex}. A phenomenological explanation of this scaling behavior based on the passive advection by turbulence was proposed in Ref.~\citep{Roche_2008}. The argument is that the vortex line density $L$ can be decomposed into two parts, $L_{||} + L_{\times}$, which behave differently. The \emph{polarized} vortex line density $L_{||}$ consists of vortex lines arranged in parallel, so as to set up a large-scale rotational flow which
follows the $k^{-5/3}$ spectrum of the turbulent normal fluid. The \emph{unpolarized} part, $L_{\times}$, is a complex tangle of vortices, so that the resulting velocity field tends to cancel out.
Because of this cancellation, the unpolarized vortex lines do not actively affect the velocity field and can be considered as a \emph{passive vector}, which is simply advected by the normal fluid. Hence, the spatial fluctuations of the $L_\times$ field 
follows the same $k^{-5/3}$ scaling as that of a passive scalar advected by the turbulence in the Obukhov-Corrsin theory~\cite{corrsin1951spectrum,obukhov1968structure,Kraichnan_1968}. By Taylor's frozen hypothesis, the frequency power spectrum has the same scaling form as the wavenumber power spectrum, hence $f^{-5/3}$. 
The fluctuations in the total line density is dominated by the $L_\times$ fluctuations, because the 
polarized vortices tend not to interact by reconnections.

\begin{figure}[t]
  \centering
  \includegraphics[width=0.45\textwidth]{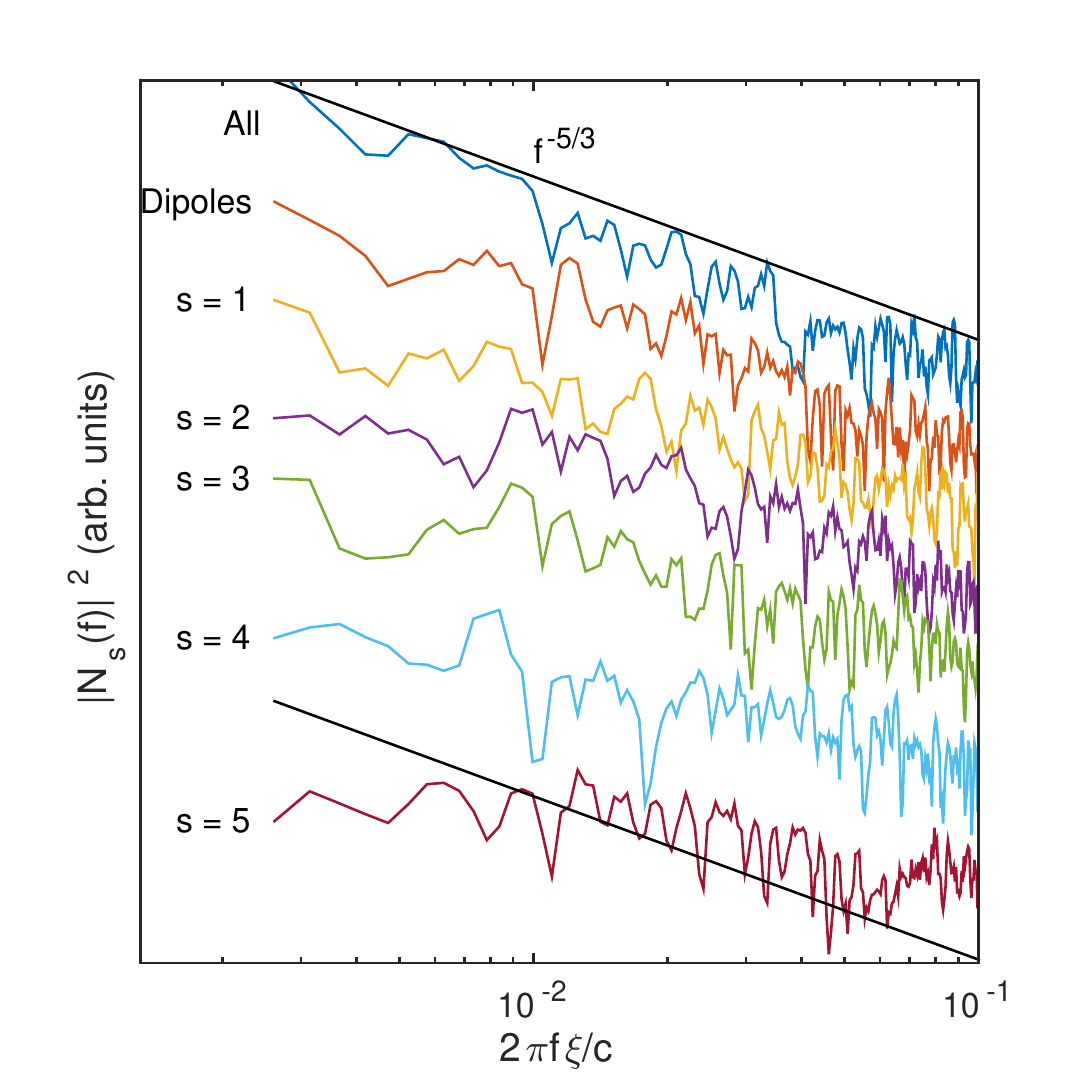}
  \caption{Power spectra for the number of vortices contained in clusters of a given size $s$, with $v_{\text{ext}} = 0.5c$. 
  The larger the cluster size, the more the power spectrum falls off from the $f^{-5/3}$ power law at low frequencies.}
  \label{fig:synspecs}
\end{figure}

We propose that a similar scenario also holds in two-dimensions and can account for the $f^{-5/3}$ power spectrum of vortex counts. The vortex density $n$ can be decomposed into the density of \emph{clustered} vortices $n_c$ and the density of 
unclustered vortices $n_u$. The clustered vortices set up a velocity field $\vec v_c$ which follows the $k^{-5/3}$ scaling, as discussed in the previous section. Their density does not fluctuate as rapidly as that of the isolated vortices, i.e. clustered vortices can be envisaged as the persistent structures. The unclustered vortices do not contribute to the energy scaling, but are passively advected by the $\vec v_c$ field. Hence the temporal fluctuations in the vortex counts is dominated by those of the isolated vortices and, using the passive scalar analogy, the power spectrum is described by a $f^{-5/3}$ on timescales corresponding to the inertial-convective regime.   

In Figure~\ref{fig:synspecs}, we show that indeed the power spectrum of isolated vortices follows a $f^{-5/3}$ scaling consistent with the passive advection model, whereas the scaling regime tends to disappear as we look at vortex clusters of increasing size.

\begin{figure}{t}
  \centering
  \includegraphics[width=0.45\textwidth]{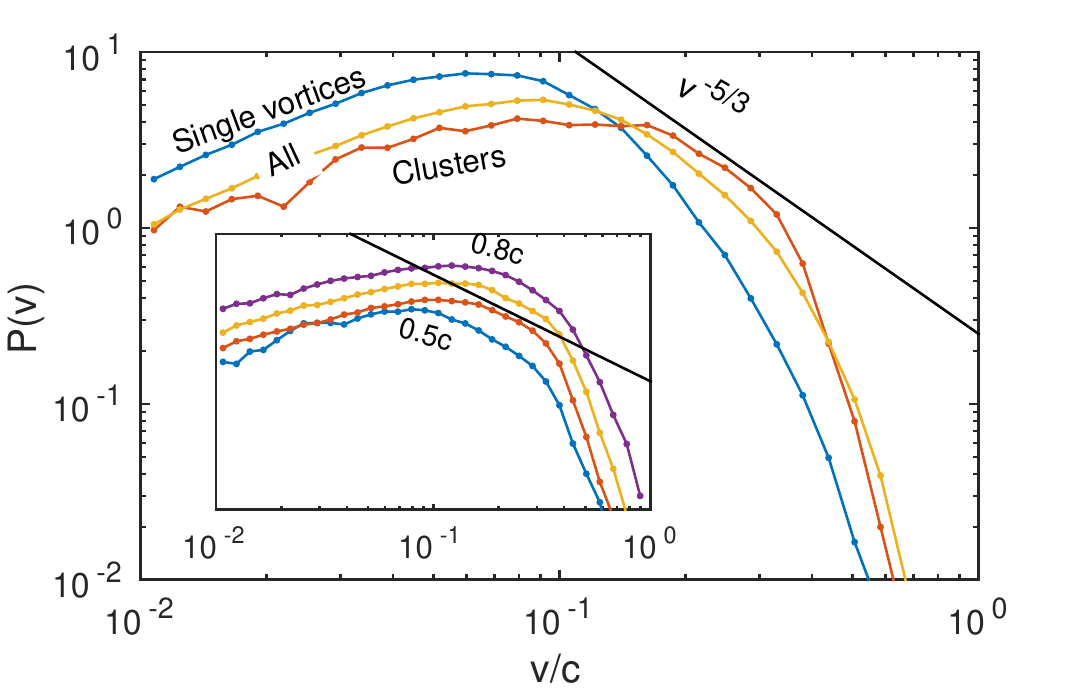}
  \caption{Vortex velocity distributions using three different sets of vortices: All vortices, single vortices, and vortices belonging to clusters of size 4.
	Inset: Vortex velocity distributions for clusters of size 3 for different stirring velocities, shifted vertically for comparison. Solid lines indicate the
	slope corresponding to a $v^{-5/3}$ scaling law.}
  \label{fig:vortpdf}
\end{figure}
\section{Vortex velocity statistics}

As discussed in Section~\ref{sec:especs}, the kinetic energy spectrum is calculated from the superfluid velocity field given by Madelung transform as $\vec v_s = \nabla \phi$, where $\phi$ is the phase of the wavefunction $\psi$. The statistics of large turbulent velocity fluctuations is however dominated by the single-vortex effects. It is known that the probability distribution of the velocity induced by a single point vortex has a power-law tail given by $p(v_s)\sim v_s^{-3}$, which can also be predicted from a simple scaling argument. Given that the velocity induced by an isolated vortex decays as $1/r$ at the distance $r$ from the vortex, then from the transformation of variables in the probabilities, $p(v_s)\ud v_s = q(r)\ud r$, it follows that $p(v_s) = q(r(v_s)) |\pd{r}{v_s}|$ where $q(r)dr$ is the probability of finding a vortex between $r$ and $r+dr$. Thus, for a uniform distribution of isolated vortices in the plane, $q(r) \propto r$, it follows that $p(v_s) \sim  v_s^{-3}$. This can also be derived from the point vortex model for configurations of $N$ uniformly distributed vortices~\citep{Chavanis_2000}. 

Because high velocity fluctuations are induced in the proximity of a vortex, and the distance between vortices is bounded below by the vortex size $\sim \xi$, the single-vortex velocity distribution dominates the high-velocity tail, at least for $v_s>c$. This is one reason that it has also been observed experimentally in quantum turbulence in superfluids~\cite{Paoletti_2008} and reproduced numerically in BEC~\cite{white2010nonclassical}. However, because the tail distribution of the superfluid velocity is dominated by contribution of single vortices, it cannot be used as a measure which can signal a turbulent cascade. 

On the other hand, the statistics of vortex velocities is an indicator of vortex clustering hence can be used as an indirect way to determine if the quantum turbulence exhibits an inverse energy cascade. As shown in Refs.~\citep{Novikov_1975,Bradley_2012}, the $k^{-5/3}$ energy spectrum is associated with vortex clusters, where vortices follow a fractal spatial distribution inside a cluster with the probability distribution $q(r)\sim r^{-1/3}$. The simple scaling argument predicts that such clustering gives rise instead to a $v^{-5/3}$ power-law tail. In a separate study~\cite{Skaugen15}, we show that the $v^{-5/3}$ also follows from the point-vortex model with a non-uniform distribution of vortices. In principle, this scaling appears in the superfluid velocity distribution at intermediate velocities $v_s<c$, but it may be difficult to observe it in practice if there are not sufficiently many vortex clusters compared to isolated vortices.

In order to sample cluster velocities more efficiently we turn to the vortex velocities. The method of locating vortices from the zeroes of the wavefunction $\psi$ also provides a way of calculating the velocity of a vortex from the time derivative and gradients of $\psi$~\citep{Angheluta_2012}. Namely, the velocity of a vortex located at position $\vec r$ is determined by the current of vortex charge, and given as
\begin{eqnarray}
v_x = \frac{1}{\mathcal{D}}\Im\left( \dot\psi\partial_y\psi^* \right), \quad
v_y = -\frac{1}{\mathcal{D}}\Im\left( \dot\psi\partial_x\psi^* \right). 
\end{eqnarray}
Numerically, we calculate the weighted average of this value across the region where $\mathcal{D}$ exceeds a threshold.

As there are only a few hundred vortices present at a given timestep, we have to gather velocity values over time in order to collect sufficient statistics for a vortex velocity histogram. The advantage, however, is that we can pick out only those vortices which belong to a cluster of a given size. This allows us to specifically sample the velocity statistics of vortices inside a clusters, and compare this to other vortices.

Such a comparison is shown in Figure~\ref{fig:vortpdf}, where we have three different velocity distributions: \emph{i)} for all vortices, \emph{ii)} for isolated vortices, and \emph{iii)} for vortices belonging to clusters of size equal to $4$ vortices. We see that the distribution of velocities corresponding to clustered vortices seems to follow a $v^{-5/3}$ scaling regime up to an ultraviolet cutoff at $v \gtrsim 0.5 c$. Like for the energy spectrum, an insufficient separation of scales limits the extent of the scaling range for $P(v)$, and additional insights are needed in order to identify the power-law. In Ref.~\cite{Skaugen15}, we showed that the scaling range for the tail distribution is controlled by the mean density of clustered vortices, and that a sufficiently low density is necessary for the onset of a scaling range. Thus, we attribute the narrow power-law tail in the numerical simulations to a high density of vortices inside clusters, which in turn is related to the fact that the vortex core size and Thomas-Fermi radius are not sufficiently far apart. 

In addition, we notice that the probability distribution of velocities sampled on isolated vortices lacks a scaling regime. This we attribute to the fact that clustered vortices and dipoles act like~\lq obstacles\rq~that prevent a uniform spread of the isolated vortices within the disk. We have checked that the $v^{-3}$-scaling appears when we redistribute the isolated vortices uniformly in the plane disregarding the presence of these~\lq obstacles\rq.      

The inset plot in Figure~\ref{fig:vortpdf} shows the distribution of velocities of clustered vortices of size $3$ for different stirring velocities. We notice that the $-5/3$ power-law tail seems to be more strongly expressed at lower stirring velocities where the contribution from clustered vortices becomes important. At higher stirring velocities, the lifespan of clustered vortices is reduced and the statistics is dominated by the isolated vortices, which do not exhibit a scaling range.

\section{Summary and Conclusions}
In summary, we have shown that the spectral energy transport in 2D quantum turbulence can be signaled from the statistics of vortex number and velocity fluctuations. This connection depends on the spatial clustering of like-signed vortices. 
To show that the inertial $k^{-5/3}$ regime of the spectral energy is due to the vortex clustering, we have studied separately the contribution of clusters and isolated vortices to the energy spectrum. 

Moreover, vortex clustering is also central to explaining the $f^{-5/3}$ scaling which we observe in the power spectrum of the vortex number fluctuations. The explanation relies on decomposing the vortex density field into clusters, which set up a prevailing velocity field, and single vortices which are passively advected by this field. Of the various signals of the inverse energy cascade, the power spectrum scaling is the most striking, as it covers a larger range of 
frequencies than the energy spectrum. One possible reason for this is that finite-size effects are less strongly expressed in the temporal domain. 
The inverse energy cascade due to vortex clustering corresponds to a particular statistical signature on the vortex velocity. We find that the clustered vortex velocity probability distribution develops a $v^{-5/3}$ power-law tail which we observed in our dGPE numerical simulations and can predict from a fractal distribution of vortices inside a vortex cluster~\cite{Skaugen15}.    

We believe that the power spectrum of vortex number fluctuations is an experimentally accessible quantity like in 3D experiments, so the predicted scaling law corresponding to the 2D inverse energy cascade can also be tested in highly oblate BECs. 

A connection between the superfluid velocity distribution and the quantum energy spectrum was also established in the hydrodynamic approximation in Ref.~\citep{Reeves_2014}, and used to study the emergence of coherent rotating structures in decaying 2D turbulence.  

We have focused on the lowest order turbulence statistics in the regime dominated by vortex dynamics. It would be interesting for the classical-quantum analogy to however go beyond the energy spectrum and study the intermittency effects. While intermittency corrections to scaling of higher order structure functions have been observed in 3D quantum superfluids~\cite{maurer1998local,barenghi2014experimental,boue2013enhancement}, this has not yet been investigated in the 2D quantum turbulence. The obvious question would be if the direct cascade is intermittent, while the inverse cascade is non-intermittent.    

\acknowledgments{Stimulating discussions with Nigel Goldenfeld, Yuri Galperin and Joakim Bergli are kindly acknowledged.}  

\bibliographystyle{apsrev4-1}
\bibliography{ref2}

\end{document}